\documentclass[twocolumn]{jetc20132e}

\usepackage{amssymb,amsmath}
\usepackage{bm}
\usepackage{epsfig}
\usepackage{color}

\def\dbar{{\mathchar'26\mkern-11mu d}}

\newcommand{\bes}{\begin{subequations}\bea}
\newcommand{\ees}{\eea\end{subequations}}
\newcommand{\be}{\begin{equation}}
\newcommand{\ee}{\end{equation}}
\newcommand{\bea}{\begin{eqnarray}}
\newcommand{\eea}{\end{eqnarray}}
\newcommand{\ba}{\begin{array}}
\newcommand{\ea}{\end{array}}

\newcommand{\bs}[1]{\boldsymbol{#1}}
\newcommand{\bx}{\boldsymbol{x}}

\title{Of dice and men. Subjective priors, gauge invariance, \\ and nonequilibrium thermodynamics}

\author{Matteo Polettini\thanks{Corresponding author.}~, Francesco Vedovato\thanks{Artwork author.}
    \affiliation{
$^\ast$ Complex Systems and Statistical Mechanics, University of Luxembourg, \\  162a avenue de la Fa\"iencerie, L-1511 Luxembourg, G. D. Luxembourg, \\E-mail: matteo.polettini@uni.lu \\}
$^{\dagger}$ Zeeburgerdijk 11b, 1093SJ Amsterdam, The Netherlands \\
Website: http://aspectacularmachine.net
}

\date {
{\small ``Ceci n'est pas une pipe'' wrote Ren\'e Magritte on what was only the {\it representation} of a pipe. Phenomena and their physical descriptions differ, and in particular the laws ruling the former might enjoy symmetries that have to be spent to attain the latter. So, inertial frames are necessary to draw numbers out of Newtonian mechanics and confront with experiment, but ultimately the laws of mechanics are independent of reference frames.  Generalizing work done in Ref. [M. Polettini, EPL {\bf 97} (2012) 30003] to continuous systems, we discuss from a foundational point of view how subjectivity in the choice of reference prior probability is a (gauge) symmetry of thermodynamics. In particular, a change of priors corresponds to a change of coordinates. Employing an approach based on the stochastic thermodynamics of continuous state-space diffusion processes, we discuss the difference between thermostatic and thermodynamic observables and show that, while the quantification of entropy depends on priors, the second law of thermodynamics is formulated in terms of invariant quantities, in particular the curvature of the thermodynamic force (gauge potential), which we calculate in a few examples of processes led by different nonequilibrium mechanisms.
}}

\begin{document}

\maketitle

\section*{INTRODUCTION}

		\begin{quote}

		{\it We can say nothing about the thing in itself, for we have eliminated
		the standpoint of knowing. A quality exists for us,
		i.e. it is measured by us. If we take away the measure, what remains
		of the quality? 
		}

		F. Nietzsche \cite{poellner}

		\end{quote}

At a first sight the varied terms appearing in the title pair as fish with bicycles. Indeed, it is our final purpose to convey that these concepts, bundled together, solve a controversy about the role of the observer in statistical mechanics, and partake to a fundamental symmetry of thermodynamics. On a less ambitious tone, objectives of this contribution are: To discuss a simple but compelling foundational aspect of nonequilibrium statistical mechanics, to extend the theory developed in Ref.~\cite{polettini} to systems with a continuous state space and, with the aid of some examples, to further back up the role of a thermodynamic curvature for the determination of the equilibrium/nonequilibrium character of steady states of diffusive systems.

 \begin{figure}[t]
\begin{center}
\includegraphics[width=260pt]{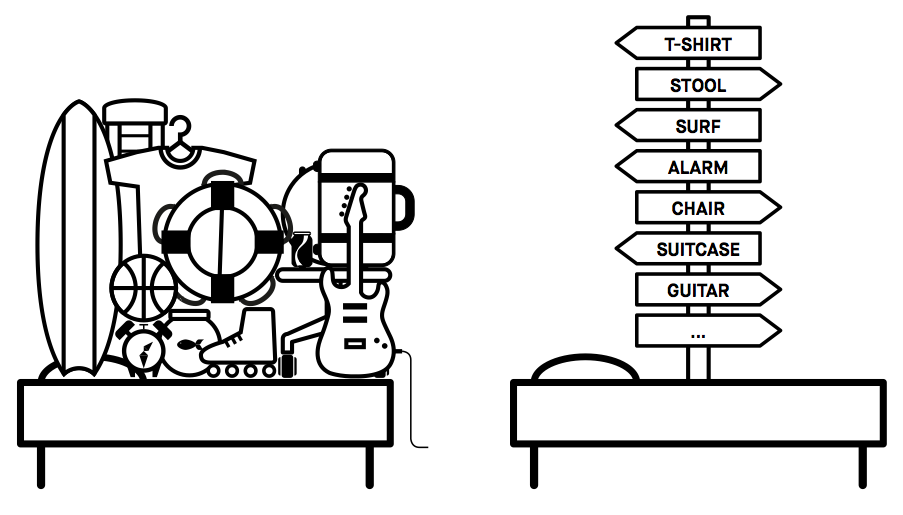}
\end{center}
\caption{A room as perceived by two observers with different prior knowledge of its state. Is there an objective criterion to quantify its disorder, given that entropy is ``missing information''?}
\label{disorder}
\end{figure}

It is renowned that statistical mechanics has been enormously successful in describing systems at thermodynamic equilibrium, bestowing a probabilistic nature on physical concepts such as that of entropy, which notoriously has two facets. The head is a state function coined in the 19th-century to put limits on the efficiency of machines. After the intuitions of Boltzmann, the tail was coined by Gibbs, and later Shannon in applications to information theory, to yield a much known but poorly understood  measure of ``disorder'' or, more precisely, of ``missing information'' \cite{farewell}.  A baffling feature of the latter acceptation is that it is prone to a certain degree of subjectivity, that only apparently does not affect its thermodynamic counterpart. We are all acquainted with the following fact   (see Fig.\ref{disorder}): As children, when mum scolded us for being messy, we would whine claiming to know exactly where our toys were. Wryly, when she tidied up we would not be able to retrieve them again. If entropy is missing information, then what is the entropy of a room? Is it the mom's barren neatness, or the child's playing strategy?

Making a huge leap upward: The Universe today presents a high degree of structure and information  (from galaxies to planets down to ourselves) due to the ordering effect of gravity, but in the far past it was a fluctuating quark-gluon plasma that cosmologists describe solely in terms of its temperature and few other parameters. Then, did entropy decrease ever since, contrary to the second law of thermodynamics?  

\begin{figure}[t]
  \centering
 \includegraphics[width=240pt]{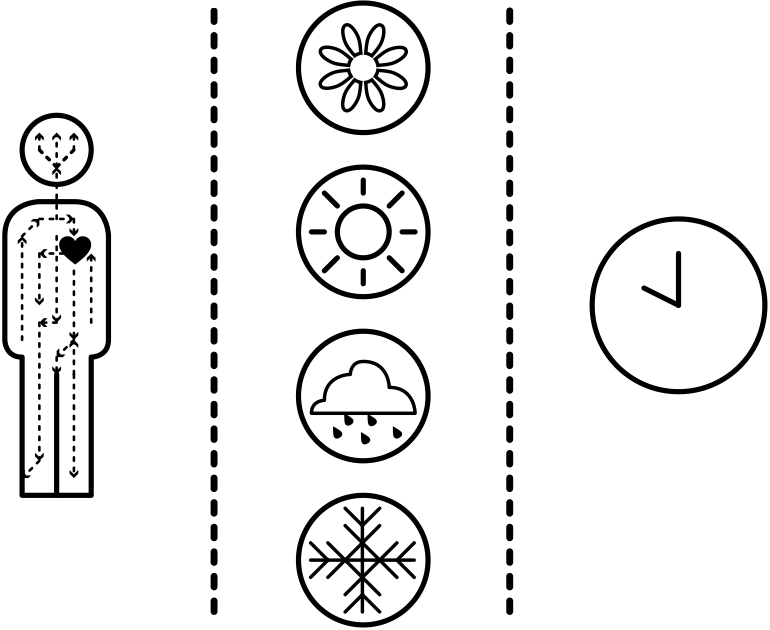}
  \caption{ \label{circulation} Currents circulate in nonequilibrium steady states.}
\end{figure}

The latter question introduces the theme of nonequilibrium processes.  Today, statistical methods encompass the response of equilibrium states to small perturbations, to embrace the sophisticated phenomenology of systems subject to nonequilibrium conditions. A special mention goes to the framework of stochastic thermodynamics \cite{stochastic}, a prominent theory that describes the thermodynamics of open systems evolving under Markovian dynamics. Nonequilibrium systems produce entropy as they evolve under the influence of thermodynamic forces towards steady states that maintain a constant heat flux towards the environment. A crucial feature of nonequilibrium steady states is the circulation of currents around loops (see Fig.\ref{circulation}). If on the one hand the {\it thermostatics} of equilibrium states is based on state functions, such as the entropy, the {\it thermodynamics} of nonequilibrium processess deals with dynamical quantities like the entropy production. The second law of thermodynamics, by many (including Einstein and Ehrenfest) considered to be the most universal law of physics, states that entropy production along a time interval $dt$ is non-negative
\be
\sigma \, dt = dS - \frac{\dbar Q}{T} \geq 0 \label{eq:second}
\ee
and that it only vanishes at equilibrium states. Here $\sigma$ is the entropy production rate and $~\dbar$ denotes an inexact differential. This law eventually provides an ``arrow of time''. But then, if entropy is a subjective quantity, will the validity of the second law and the direction of time depend on the observer?

Similar dreaded outlooks led to criticisms about the actual relevance of the information-theoretic approach to thermodynamics, as pioneered by Jaynes \cite{jaynes}. For example Ref.~\cite{lebowitzmaes} is a funny skeptical fable about an obtuse physicist questioning an omniscient angel about who is the ``right'' observer. At the 11th Granada Seminar \cite{granada} Lebowitz prodded the scarce informationist supporters that an observer who's no Ph.D. might threaten the fate of physical laws. 

Our apology of the informationist approach supports the following point of view. A change of observer is analogous to a change of reference frame in mechanics, or of coordinate system in general relativity: It does not alter the process, but it does alter the representation of the process. While it is always necessary to choose an observer to actually do physics (and this choice can be done in a more or less awkward way\footnote{For example, it doesn't make much sense to describe the ballistics of a rocket on the Earth using the rest frame of a merry-go-round on the Moon, although in principle it is possible.}), it is necessary that the fundamental laws remain invariant. For this reason, while thermostatic quantities like the entropy can change, it is necessary that the entropy production rate involved in the second law does not. In other words, changes of subjective observer must be a symmetry of nonequilibrium thermodynamics. In particular, as we will see, it can be implemented as a so-called {\it gauge} symmetry, that is, a symmetry that acts locally, from point to point, rather than globally (like a change of units does).

\section*{PLAYING DICE AND SITTING ON STOVES}

Dice are a common tool for intuitive probabilistic thinking, so much that Einstein declared that ``God doesn't play dice'', objecting the probabilistic nature of quantum mechanics (but he would also say that ``physical concepts are free creations of the human mind, and are not uniquely determined by the external world''). Sticking to classical statistical mechanics, we don't dare challenge the divine intentions, and rather consider different human observers rolling dice.

Given the common-sense symmetry of a die with respect to its bouncing on a table, one is compelled to assign prior $1/6$ to all faces. This seems such an obvious choice, that it is hard to admit that it is just as subjective as any other. This prior is related to a measure of a die's properties, made by an observer who has a sufficiently complete perspective on the die with respect to the process being performed and to uncountable previous experiences of seeing and hearing of similar dice rolling. It is nevertheless legit to color a die's faces one red, one green, one blue and all others yellow, and have an heavily myopic person examine it from a distance, so that he can only resolve spots of color. He will have sufficient reason to believe that the system is a tetrahedron. According to his most plausible probability assignment, the die'sfaces would have probability $(1/4,1/4,1/4,1/12,1/12,1/12)$ to show up. From his perspective, this makes perfectly sense.

\begin{figure}[t]
\begin{center}
\includegraphics[width=260pt]{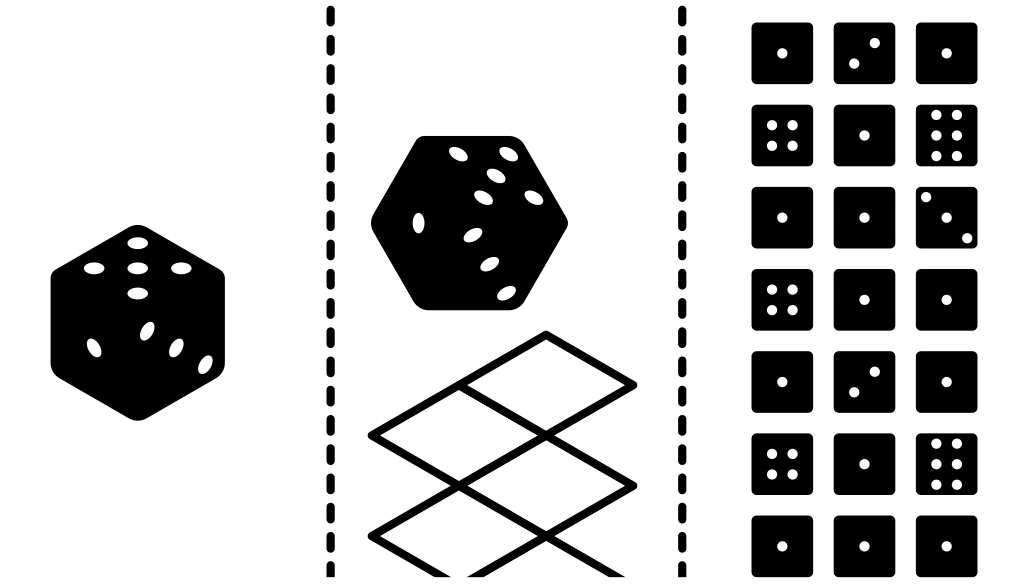}
\end{center}
\caption{A die and the process of throwing a die suspect of being loaded, given the biased set of outputs on the right-hand side.}
\label{dice}
\end{figure}

Suppose that ourselves and the myopic person play dice, correcting our prior according to the output. If, by the law of large numbers, the die's faces come out approximately $1/6$ of the times, we the seeing will gather no further information, since we had a fully satisfactory estimation of the die's entropy in the beginning. The myopic, who had a worse estimation, will gain further information. By the paradigm that information is physical, we can say that our measure of the initial entropy and of the forthcoming heat fluxes differed from his. Nevertheless, the process occurred in the same way. This {\it gedankenexperiment} can provoke an objection: The half-blind person is in a chronicle state of ignorance and he doesn't see the ``truth". But so are we with respect to a loaded die, in which case we either know technical details about how it was loaded and then formulate a reasonable prior, or else we can just make an hypothesis and then correct it according to the output, as in Fig.~\ref{dice}.

\begin{figure}
  \centering
 \includegraphics[width=260pt]{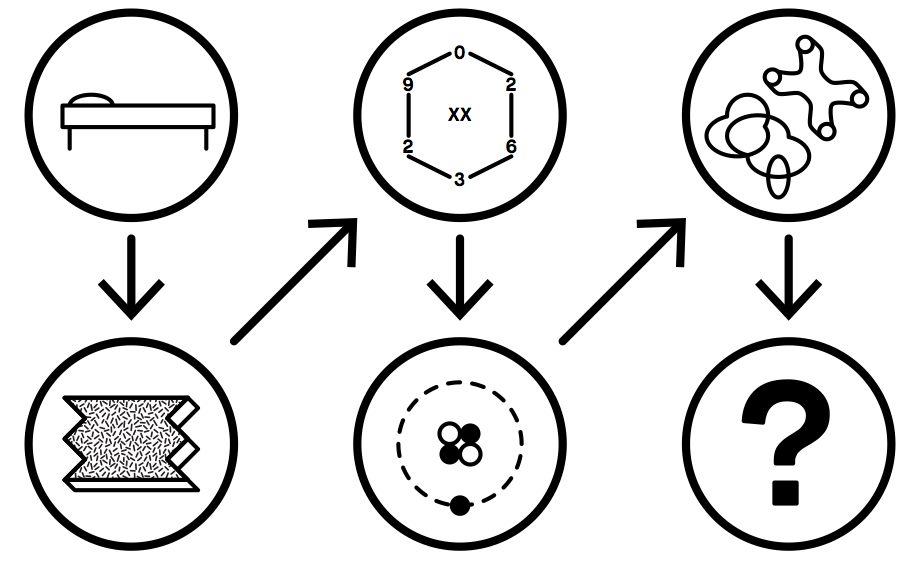}
  \caption{ \label{micro} Is there an ultimate temperature of a bed?}
\end{figure}

Let's now move to physical quantities, in particular temperature. Consider a hot stove, and let's ask:
\begin{quote}
What \textit{is} the stove's temperature?
\end{quote}
Taking a na\"\i{}ve materialistic approach, one could say a stove is a stove and it has the temperature it has. We can sit on it and perceive it. But things are more sophisticated. We ourselves are thermometers. Every-day thermometers interact with certain degrees of freedom of the system, say, the electromagnetic forces of the outer non-shielded electrons of a solid's molecules, but essentially do not interact strongly with the nuclei, electroweakly with neutrinos, they do not exchange gluons, massive vector bosons, gravitons, strings, quanta of space-time and whatever is beyond our present state of knowledge. But one can in principle put the stove in an accelerator and get a much more accurate estimate of the temperature. So, the answer to the above question depends on the coarse graining of ``reality'' that physical apparatuses always entail. We leave aside the question whether there exists an ultimate temperature (see Fig.\ref{micro}).

A different question that we can pose is:
\begin{quote}
What \textit{happens} when a thermometer is put on the stove?
\end{quote}
Italics was used to emphasize their different natures: The first questions an  essential property of the system, while the second concerns a process which occurs when two physical systems are put in relation. Much like in the Zen koan ``What is the sound of one hand?'', we claim that the first question is unphysical (but, rather, metaphysical), while the second is physical, and giving an answer necessarily calls into play observers. If we sit  on the stove, we will get burnt in a few seconds, again according to an Einstein's estimate, independently of what we think of it.

The punchline is: We should not worry about subjectivity of physical quantities, just as much as we don't deem it necessary to wear another coat when we express $75.2^\circ$ Fahrenheit as $24^\circ$ Celsius. We should rather make sure that fundamental physical laws are independent of this choice of reference frame. This is the main objective of this paper.

\section*{SWITCHING PRIORS AND COORDINATE FRAMES}

Let $\bx \in X$ be a generic state of a system, labeling some microscopic degrees of freedom (for example, positions and momenta of the molecules of a gas, spin states of a ferromagnet etc.). For simplicity we suppose that the state space has finite volume normalized to unity, $\int_X d\bx = 1$. Statistical descriptions of the system assign a probability density $p(\bx)$ to microstates. For example when the mechanisms involved in a physical process involve exchange of a form of energy with a single heat bath at temperature $T$ the most plausible distribution compatible with an observed average value of the  energy $\langle E \rangle$, assuming that the underlying microstates are equiprobable, is Gibbs's canonical distribution $p_{\mathrm{Gibbs}}(\bx) =\exp - [E(\bx) - F]/(k_B T)$, with $F$ Helmholtz's free energy and $k_B$ Boltzmann's constant, that we set to unity in the following. Notice that our prudent wording emphasized that the choice of probability density is congenial to a particular process, and that there is a choice of prior involved.

The Gibbs-Shannon (differential) entropy 
\be
S = - \int_X d \bx  \, p(\bx) \ln p(\bx) \label{eq:S1}
\ee
is a measure of the missing information of the probability distribution with respect to a state of perfect knowledge. It 
vanishes when $p(\bx) = \delta(\bx-\bar{\bx})$, and it is maximum when the distribution is uniform. As announced, there is a correspondence between statistical and physical entropies, as one can appreciate by plugging the canonical distribution to recover the well-known expression between equilibrium thermodynamic potentials $TS = \langle E \rangle - A$. 

In regard to probability densities, an important mathematical detail that we need to point out is that they are obtained by taking the so-called Radon-Nikodym derivative of a probability measure $P$ with respect to another\footnote{We assume that all probability measures are absolutely continuous with respect to the uniform distribution.} that we call the prior $P_{pr}$,
\be
p = \frac{dP}{dP_{pr}}.
\ee
In Eq.~(\ref{eq:S1}) the prior is implied to be the uniform normalized distribution over microstates, $dP_{pr} = d\bx$. Hence the definition of entropy always pivots on a prior. Also, the canonical distribution can be obtained as the maximum entropy distribution compatible with a measured value of the average energy $\langle E \rangle$, assuming the uniform prior over microstates, viz. starting with the microcanonial ensemble.

Let us rewrite entropy in terms of probability measures: 
\be
S = - \int_X dP  \ln  \frac{dP }{d\bx}. \label{eq:S2}
\ee
Our point of view in this paper is that the choice of uniform prior is just as subjective as any other choice, and that changes of priors
\be
d\bx \to dP_{pr}',
\ee
{\it at fixed probability measure $P$} (that is, at fixed macrostate) are legitimate and need to be coped with in a consistent manner. Under such a transformation we obtain
\be
S' = - \int_X dP   \ln \frac{dP}{dP_{pr}'} = S + \left\langle \ln \frac{dP_{pr}' }{d\bx}\right\rangle, \label{eq:St}
\ee
where the average is taken with respect to $P$. 
Entropy is not an invariant object, as it gains an additional term. It is also well known that entropy is not invariant under orientation-preserving coordinate transformations $\bx \mapsto  \bx'(\bx)$, with Jacobian
\be
\Lambda = \det  \left( \frac{\partial \bx'}{\partial \bx} \right) > 0. 
\ee
Being the probability measure a volume form, viz. $dP = dP'$ so as to preserve probabilities of events, and since the volume element transforms according to $d \bx' = \Lambda  d \bx$, one finds the transformation law for the probability density $p' = \Lambda^{-1} \, p$. Plugging into Eq.~(\ref{eq:S1}),  under a change of coordinates the entropy gains an inhomogeneous term
\bea
S' = - \int_{\bx'(X)} dP'   \ln \frac{dP'}{d\bx'}  = S + \langle \ln \Lambda \rangle. \label{eq:Str}
\eea
Notice that a volume-preserving transformation with $\Lambda = 1$  preserves the entropy. This is the case for canonical transformations in Hamiltonian mechanics and for the Hamiltonian flux (indeed it is inappropriate to say that ``entropy of an isolated system cannot decrease'', since it is a constant of motion). However, volume-preserving transformations are too restrictive. For example, in the approach to nonequilibrium thermodynamics based on dynamical systems and Gaussian thermostats, evolution does not preserve the phase space volume \cite{ruelle}. 

The analogy of Eq.~(\ref{eq:Str}) with Eq.~(\ref{eq:St}) suggests that every change of prior can be achieved by a change of coordinates with $dP_{pr}' = d\bx'$. In fact, inspecting Eq.~(\ref{eq:S2}) we realize that a coordinate transformation maintains $dP$ but replaces $d\bx$ with a new prior. In the new coordinates $d\bx'$ is the uniform measure while $d\bx$ is not anymore. A coordinate transformation realizes a change of the relevant degrees of freedom that are supposed to be equiprobable. This is well illustrated by this riddle: Picking a number $\bx$ between $1$ and $10$ at random, the probability that it is smaller than $5$ is $1/2$, while picking $\bx'$ at random between $1$ and $100$, the probability that it is smaller than $25$ is $1/4$. How is it possible that picking either a number or its square aren't equally likely? The solution is to recognize that different choices of prior were made in the two cases, and that the uniform prior (``at random'') in one set of coordinates is not the uniform prior in the other set of squared coordinates.

To some authors, non-invariance of the entropy sounds as a dazzling puzzle that discredits its utility. The italian-speaking readers will also find a discussion on this tone in Ref. \cite{vulpiani}, which comes to the conclusion that Jaynes's MAXENT reasoning is circular. Our point of view is that there is not a preferred set of coordinates, and while the determination of thermostatic quantities does depend on coordinates, thermodynamics should not. Hence the entropy must be complemented with another quantity that grants the overall invariance of thermodynamics.

\section*{ENTROPY PRODUCTION RATE}

We now suppose that the system is in contact with an environment that determines a Markovian evolution of its probability density, dictated by the Fokker-Planck equation
\bea
\dot{p} = - \nabla \cdot \big( p \, \bs{A} - T \nabla p \big) = - \nabla \cdot \bs{J},  \label{eq:fpA} 
\eea
with $\bs{A}$ a thermodynamic force and $T$ an environmental temperature. The dot derivative is with respect to time. On the right-hand side we put it in the form of a continuity equation in terms of a probability density current $\bs{J}$. Under mild assumptions the Fokker-Planck equation evolves towards a unique steady state $p^\ast$, at which the current has no sinks and sources, $\nabla \bs{J}^\ast = 0$ (the asterisk will mark the steady value of any observable). A steady state is in equilibrium when the steady state currents vanish, while nonequilibrium steady states are characterized by nonvanishing currents that circulate in the system's state space.

Since this equation regulates the dynamics of the process, we assume it to be invariant under change of priors/coordinates. Unfortunately, a fully satisfactory treatment would require more advanced tools from the theory of differential forms, including the introduction of a metric. Ref. \cite{polettini3} contains further details. The current transforms like a vector density. A suitable transformation law for the divergence grants that $\nabla \cdot \bs{J}$ is a scalar density. Within the current, under a change of coordinates the gradient of the probability density develops an inhomogeneous term that must be reabsorbed by the thermodynamic force. Hence overall invariance enforces the transformation laws
\be
p' = \Lambda^{-1} p, \qquad \bs{A}' =  \frac{\partial \bx}{\partial \bx'} \left( \bs{A} - T \nabla \log \Lambda \right).
\ee
This pair of equations is reminiscent of gauge transformations as encountered in (quantum) field theory, with $p$ playing the role of the wave function, $\bs{A}$ that of gauge potential, and $T$ that of coupling constant. As a technical note, the gauge group in this case is the noncompact group of real positive numbers under multiplication, whose elements are the Jacobians of orientation-preserving diffeomorphisms.

One can now use the transformation law for the gauge potential to counterbalance the inhomogeneous term developed by the entropy. To this purpose we prefer to consider the rate of entropy production  $\dot{S}$. Its transformation law is
\be
\dot{S}' = \dot{S} + \int \bs{J} \cdot  \nabla \log \Lambda \, d \bx . 
\ee
It can then be seen\footnote{Notice that the scalar product in this expression denotes the presence of a metric, which also has to transform properly altogether.} that the following {\it heat flux rate}
\be
\frac{\dbar Q}{dt} = - \int \bs{J} \cdot \bs{A} \, d \bx 
\ee
has exactly the same transformation law as $\dot{S}$ but for a factor $1/T$, so that the {\it entropy production rate}
\be
\sigma = \dot{S} - ~\frac{1}{T} \frac{\dbar Q}{dt}  \qquad [~\dbar S_i = dS - ~\dbar S_e]
\ee
is indeed an invariant quantity (between square parenthesis we reported the analogous decomposition of the entropy production as found in older thermodynamics textbooks \cite{glans}). This expression for the entropy production rate is well-known in the stochastic thermodynamics literature \cite{stochastic}. It can easily be proven to be positive, which is a statement of the second law, Eq.~(\ref{eq:second}). By construction $\sigma$ is invariant under gauge transformations, hence the second law holds in all coordinates/with respect to all choices of prior.

\section*{CURVATURE AND STEADY STATES}

In general, $\bs{A}$ is not a conservative force, that is, it is not a gradient. Then $\sigma$ is not the total time derivative of a state function, from which it follows that $~\dbar Q^\ast/dt \neq 0$ and the steady state value of the entropy production does not vanish: Nonequilibrium steady states maintain a constant entropy production as currents circulate. Instead when $\bs{A} = - \nabla \Phi$, which we refer to as the condition of {\it detailed balance}, then $\sigma \, dt$ is an exact differential and it vanishes at the steady state.

The characterization of nonequilibrium can be elegantly made in terms of the curvature of the gauge potential. Curvature is a crucially important quantity in gauge theories. For example, in the field-theoretic formulation of electromagnetism, the curvature tensor contains the components of the electric and magnetic fields and entails the peculiar differential form of Maxwell's equations. In this section we discuss thermodynamic curvature and express the steady state entropy production in terms of it. The main messages are that one should consider several different contributions to the curvature, and that a vanishing curvature is intimately related to the occurrence of equilibrium steady states.

\subsection*{Nonconservative force} 

The curvature associated to the non-conservative force $\bs{A}$ in Eq.~(\ref{eq:fpA}) is defined as
\be
F_{ab} =  \frac{\partial A_b}{\partial x^a}   -  \frac{\partial A_a}{\partial x^b}, \label{eq:tenscur}
\ee 
Here $a,b$ denote vector and tensor components. Notice that it transforms like a 2-tensor under coordinate transformations and it does not gain inhomogeneous terms, hence it is gauge invariant. When the force is a gradient, then the curvature vanishes. The converse is not generally true, as there might be topological contributions to the entropy production, such as isotropic flows around the fundamental cycles of a torus. Such topological contributions have been studied by several authors in Ref. \cite{qians}. Here we discard them.

A theorem by Hodge implies that the current can always be decomposed as (letting $\delta^{ab}$ be the Kroenecker's delta)
\be
J^a = \sum_b \frac{\partial}{\partial x^b} \left(\Theta^{ab}-    \Psi \, \delta^{ab}\right) + \Omega^a, \label{eq:decom}
\ee
where $\Psi$ is a scalar potential, $\Theta$ is a skew-symmetric tensor, and $\Omega$ is an harmonic vector, such that $\Delta \Omega = 0$. The latter term is the one that provides topological contributions, and we discard it. Taking the gradient, since $\Theta$ is skew-symmetric, we find that
\be 
- \nabla \cdot \bs{J} = \Delta \Psi = \dot{p}.
\ee
In particular at a steady state one has $\Delta \Psi^\ast = 0$. It is known that on a compact manifold without boundary the only harmonic scalar is a constant function $\Psi^\ast = const.$, hence we find that only the term in $\Theta^\ast$ survives in the expression for the steady current. Plugging into the entropy production rate and integrating by parts we obtain the steady state value
\be
\sigma^\ast =  \frac{1}{2T} \sum_{a,b} \int F_{ab} \Theta^{\ast\,ab}  d\bx.
\ee
This result can be seen as a continuous version of a decomposition by Schnakenberg of the steady entropy production rate in terms of fundamental cycles of a graph \cite{schnak}.

\subsection*{Competition between heat reservoirs}

We now consider a Brownian particle interacting with two baths at different temperature. Its stochastic motion is described by the differential equation
\be
\dot{\bx} = - \nabla \Phi_1 + \sqrt{2T_1}\, \bs{\zeta}_1  - \nabla \Phi_2 + \sqrt{2T_2} \, \bs{\zeta}_2,
\ee
where, as usual, $\bs{\zeta}_1$ and $\bs{\zeta}_2$ are uncorrelated sources of white noise, $\langle \zeta^a_i(t) \zeta^b_j(t') \rangle = \delta^{ab}\delta_{ij} \delta(t-t')$, $i,j \in (1,2)$. The key aspect regarding this system is that, although the two forces are gradients (condition of {\it local detailed balance} \cite{ldb}), meaning that detaching either bath returns equilibrium, competition between the two baths to impose their own equilibrium gives rise to nonequilibrium character, and further curvature terms.

The Fokker-Planck equation for this system reads
\be
\dot{p} = - \nabla \cdot \bs{J}_1 - \nabla \cdot \bs{J}_2 =   \nabla \cdot \big(p\nabla \Phi_1 + T_1 \nabla p  + p \nabla \Phi_2 + T_2 \nabla p \big). 
\ee
It is known that a proper description of the thermodynamics of the system requires to resolve the two mechanisms, otherwise one would systematically underestimate the entropy production  \cite{ldb}. For this reason we distinguished two currents that flow in parallel and we resolve two driving forces $\bs{A}_i = -\nabla \Phi_i$. The steady state can be easily computed giving $p^\ast \propto \exp - (\Phi_1 + \Phi_2)/(T_1+T_2)$, and while the total steady current $\bs{J}_1^\ast + \bs{J}_2^\ast$ vanishes, one finds that the individual currents do not:
\be
\bs{J}^\ast_1 = - \bs{J}^\ast_2 = \frac{p^\ast}{T_1 + T_2} \left( T_2 \nabla \Phi_1 - T_1 \nabla \Phi_2 \right). 
\ee
Again, we can perform the Hodge decomposition of $\bs{J}^\ast_1 $ as in Eq.~(\ref{eq:decom}) to obtain 
\be
\sigma^\ast =  \int \left (\frac{\bs{A}_1 \cdot \bs{J}^\ast_1}{T_1} + \frac{\bs{A}_2 \cdot \bs{J}^\ast_2}{T_2}\right)  d\bx  = \int \left(\frac{\Delta \Phi_2}{T_2} - \frac{\Delta \Phi_1}{T_1}  \right) \Psi^\ast_1 \,  d\bx .
\ee
Notice that $\Psi^\ast_1$ in this case does not vanish, since $\nabla  \bs{J}^\ast_1 \neq 0$. The above formula shows that when several baths compete one shall also consider contributions from the cross scalar curvature $\Delta \Phi_2 / T_2 - \Delta \Phi_1/ T_1$ even if local detailed balance holds.

\subsection*{Blowtorch}

Systems subject to a temperature gradient undergo the so-called blowtorch effect, first described by Landauer \cite{blowtorch}: Even if the thermodynamic force is conservative, the varying temperature profile $T = T(\bx)$ might turn it into a nonequilibrium driving force. 
The steady entropy production now reads
\be
\sigma^\ast = - \int \frac{\nabla \Phi \cdot \bs{J}^\ast}{T} d\bx =  \frac{1}{2} \sum_{a,b} \int F^{(T)}_{ab}  \Theta^{\ast\, ab}  d\bx,
\ee
where again we employed Hodge's decomposition and introduced an additional curvature term given by
\be
F^{(T)}_{ab} =  \frac{\partial}{\partial x^a}  \Phi  \frac{\partial}{\partial x^b}  \frac{1}{T} -   \frac{\partial}{\partial x^b}  \Phi  \frac{\partial}{\partial x^a}  \frac{1}{T}.
\ee
The thermodynamics force $\nabla (1/T)$ appears. 
The blowtorch effect vanishes when $\Phi(\bx) = \Phi(T(\bx))$. 
It can be shown that this curvature term is the infinitesimal version of a cycle $\oint d\Phi/T$ when the integral is performed along a small square cycle with sides in the $a$-th and $b$-th directions. The latter expression is reminiscent of Clausius's expression for the entropy along a closed individual realization of a process (rather than of an ensemble). In a recent paper one of the authors \cite{polettini2} interpreted temperature gradients as a deformation of the metric of space. 

\section*{CONCLUSIONS}

		\begin{quote}
		
		{\it You never oughta drink water when it ain't runnin'.}

		J. Steinbeck, {\it Of Mice and Men} 

		\end{quote}

In this paper we faced the problem of subjectivity of information-theoretic entropy under a change of reference prior probability, and turned it into an opportunity for a symmetry principle of nonequilibrium thermodynamics. We argued that the physical counterpart of a reference prior choice is the inherent coarse-graining that any description of a physical system entails. We observed that\ while thermostatic quantities pertaining to fixed states such as the entropy need not be invariant, fundamental laws pertaining to processes like the second law of thermodynamics must be independent of the observer. We then formulated transformation properties as so-called gauge transformations and built an appropriate gauge-invariant entropy production, returning a well known expression in the framework of the stochastic thermodynamics of diffusion processes. At a nonequilibrium steady state, the entropy production can be expressed in terms of the curvature of the nonequilibrium force (gauge potential). Several contributions coming from different nonequilibrium mechanisms have been described.

From a slightly more philosophical perspective, we supported the informationist approach to statistical mechanics, rejecting imputations of solipsism (``reality doesn't exist'') or cognitive-relativism (``reality depends on the observer''), but rather arguing that it is a very laic and prudent point of view that purports that the only physically meaningful concepts are those that can be measured, and measures require observers. In other words, there are no omniscent angels in physics. Nevertheless, we showed that physical laws such as the second law of thermodynamics are ultimately be independent of the observer.

\begin{acknowledgment}
Artwork by designer Francesco Vedovato. Discussion with A. Vulpiani and A. Montakhab is at the origin of this work. The research was in part supported by the National Research Fund Luxembourg in the frame of project FNR/A11/02.
\end{acknowledgment}

\begin{nomenclature}
\entry{}{\hspace{-0.5cm} All quantities can be considered to be dimensionless.}
\entry{}{}
\entry{$\bs{A}$}{Thermodynamic force / gauge potential}
\entry{$F_{ab}$, $F^{(T)}_{ab}$}{Curvature, blowtorch curvature}
\entry{$\bs{J}$}{Probability density current}
\entry{$p$}{Probability density}
\entry{$P$, $P_{pr}$}{Probability measure, prior probability measure}
\entry{$Q$, $~\dbar Q/ dt$}{Heat, heat flux rate}
\entry{$S$}{Gibbs-Shannon entropy}
\entry{$t$}{Time}
\entry{$T$}{Temperature}
\entry{$\bx$, $X$}{Microstate, state space}
\entry{$\Lambda$}{Jacobian of a coordinate transformation}
\entry{$\sigma$}{Entropy production rate}
\entry{$\bs{\zeta}$}{White noise}
\entry{$\Phi$}{Scalar potential for the force}
\entry{$\Psi$, $\Theta$, $\Omega$}{Scalar, tensor and vector current terms}
\end{nomenclature}

\endmytext

\end{document}